\documentstyle[prd,aps]{revtex}
\begin{document}
\def\eq{\begin{equation}}
\def\en{\end{equation}}
\newcommand{\beqn}{\begin{eqnarray}}
\newcommand{\eeqn}{\end{eqnarray}}
\def\part{\partial}
\def\pprime{{\prime \prime}}
\def\part{\partial}
\def\partr{\partial_{r_c}}
\def\query{\marginpar{?}}
\def\sf{(\sigma_f-\sigma)}
\def\sfd{(\sigma_f-\sigma_m - \delta)}
\def\hst{\frac{H \sigma}{2}}
\def\jarg{\frac{\sqrt{9- \Lambda^2}}{2}}

\preprint{UIUC-THC-98/4;astro-ph/9804289}
\title{Fluctuations in the Hawking Turok model}
\author{J. D. Cohn}
\vskip 1cm
\address{Departments of Physics and Astronomy,
University of Illinois \\
at Urbana-Champaign, IL 61801 \\
{\rm jdc@uiuc.edu} \\}
\date{April 1998}
\maketitle

\begin{abstract}
Scalar and gravity wave subcurvature fluctuations are calculated for
a background approximating the Hawking and Turok
open universe model.  The gravity wave cosmic microwave background
contribution is finite and it appears
that a normalizable scalar supercurvature mode is possible in some
regions of parameter space.
\end{abstract}
\pacs{Preprint UIUC-98/4; \hskip 0.5 cm
astro-ph/9804289}
\vskip 2pc
\section{Introduction}
The Hartle-Hawking wavefunction\cite{hh} has been suggested to describe
`tunneling from nothing' and has recently been proposed as a way
to produce an open universe\cite{ht}.
The interpretation of this mechanism, the most likely
value of $\Omega$ and the singular nature of the background
are all under study and criticism at the moment (with varying interpretations,
counterexamples and extensions in
\cite{linde,unruh,lin-b,alex,garriga,gfive,interps}).
In the following, the subcurvature scalar and (`electric parity') gravity wave
fluctuations are calculated for a model of this sort,
using the approximate form for the
background given in the papers above.  The `electric parity' gravity waves
contribute to the cosmic microwave background (CMB)
while the `magnetic parity' ones do not.

It will be assumed that the singularity does not cause
a difficulty aside from imposing boundary conditions on
basis functions to be regular in its presence.  This requirement
is discussed in \cite{garriga}. The time dependence will
be taken to be the same, at `tunneling time', as that of the Bunch-Davies
vacuum, for reasons described in section two below.
It will be assumed that
the value of $\Omega$ is appropriate.

One reason this calculation might be of interest is that for a Bunch Davies
vacuum in an open universe, the contribution of
gravity waves to the cosmic microwave background is divergent\cite{all-cal}.
In the model considered here, the subcurvature gravity wave contribution
is found
to be finite.
The background metric and field are described in the rest of this section.
In the next section the scalar perturbations are considered, followed by
a section on supercurvature modes and then gravity waves.

The instanton describes several regions of spacetime (see \cite{ht}
for further details).
The Euclidean (tunneling) part of the wavefunction is
described by a metric
\eq
ds^2 = d \sigma^2 + b^2(\sigma)(d \psi^2 + \sin^2 \psi d \Omega_2^2)
\en
At the `tunneling time' $\psi = \pi/2$, one can continue this to
the
Lorentzian region by
\eq
\psi = \pi/2 + i \tau
\en
The metric then becomes
\eq
ds^2 = d \sigma^2 + b^2(\sigma)(-d\tau^2 + \cosh^2\tau d\Omega^2_2)
\en
As $\sigma$ is a spatial coordinate, this spacetime is not
of FRW form.  Here, the parameter $\sigma$ runs between
$\sigma_f \ge \sigma \ge 0$, where $b(0) = b(\sigma_f) = 0$.

The open universe is the forward light cone of the point $\sigma=0$
with $\sigma = it, \tau = i \pi/2 + \chi $ and $a(t) = - i b(it)$.
The metric in this region is
\eq
ds^2 = - dt^2 + a(t)^2(d \chi^2 + \sinh^2 \chi d\Omega^2_2)
\en

There is also a closed universe obtained by
continuation at the maximum radius $\sigma = \sigma_{max}$
with
\eq
ds^2 = - dT^2 + b^2(T)(d \psi^2 + \sin^2 \psi d\Omega_2^2)
\en
That is $0 \le \sigma \le \sigma_{max}$ for the Euclidean region and
$\sigma = \sigma_{max} + iT$ in the Lorentzian region.

The action is
\eq
S_E = \int d^4 x \sqrt{g} (-V(\phi) + \frac{1}{16} F^2)
+ \frac{1}{8 \pi G} \int d^3 x \sqrt{h} K \; .
\en
The boundary term,
$\frac{1}{8 \pi G} \int d^3 x \sqrt{h}K
= -\frac{1}{8 \pi G}(b^3)^\prime \int d\Omega^3 \; $, comes in
as there is a small area excised around $\sigma= \sigma_f$,
and the equations of motion have been used.

We will use a linear potential here, $V(\phi) = V_{,\phi} \phi$.
The equations of motion in terms of $\sigma$ are
\eq
\begin{array}{l}
\phi^\pprime(\sigma) + 3 \frac{b^\prime(\sigma)}{b(\sigma)}\phi^\prime =
V_{,\phi}
\\
(b^\prime(\sigma))^2 - 1 = \frac{\kappa b(\sigma)^2}{3}
(\frac{\phi^{\prime \, 2}(\sigma)}{2} - V_F(\phi(\sigma)))
\end{array}
\label{eom}
\en
Here $V_F = V(\phi)+  F^2/48$, and $F$ is a constant \cite{cos}.

The instanton background solving these equations is, to leading order
\cite{ht},
\eq
\begin{array}{lll}
b_R(\sigma) \sim \sigma \; \; \; \; \; \;
& \phi_R(\sigma) \sim \phi_0 + O(\sigma^2) & \sigma \sim 0 \\
\nonumber
b_L(\sigma) \sim C(\sigma_f-\sigma)^{1/3} & \; \;
\phi_L(\sigma) \sim -\sqrt{\frac{2}{3\kappa}}\ln \sf & \;
\; \;  \sigma \sim \sigma_f
\label{asymp}
\end{array}
\en
with $\kappa = 8 \pi G$.

A background solution for $\phi,b$
for all $\sigma_f \ge \sigma \ge 0$ will be obtained
by extending the background metric near both endpoints and
connecting them near $\sigma \sim \sigma_f$.
Extend the background metric near $\sigma \sim 0$,
by $b(\sigma) =H^{-1} \sin H \sigma$.   The
background field configuration $\phi_b$ then obeys \cite{bgt}
\eq
\part_\sigma \phi_{b,R}(\sigma) =
\frac{V_{,\phi}}{6 H} \frac{\sin \frac{H \sigma}{2}}
{(\cos \frac{H \sigma}{2})^3} (2 + \cos H \sigma) \; ,
\en
where $\part_\sigma \phi_{b,R} (\sigma)\vert_{\sigma = 0} = 0$ has been used.

For $\sigma \sim \sigma_f$, to leading order,
$b_L \sim C(\sigma_f - \sigma)^{1/3}$ as above.  To go to higher order in
$\sf$,
write
\eq
b_L(\sigma) = C \sf^{1/3} (1 + A \sf^{4/3}) \; .
\en
Dimensionally, $[C] \sim \sigma^{2/3}, \; [A] \sim \sigma^{-4/3}$.
The power of the additional term has been chosen to cancel
the curvature in the equations of motion (\ref{eom}).
Then
\eq
\part_\sigma b_L(\sigma) = -\frac{C}{3} \sf^{-2/3}(1 + 5 A \sf^{4/3}) \; .
\en
so that the first equation of motion in (\ref{eom}) then gives
\eq
\part_\sigma \phi_{b,L} =
\sqrt{\frac{2}{3 \kappa}}
(\frac{1}{\sf} - 3 A \sf^{1/3}) - V_{,\phi} \sf/2 + O(\sf^{5/3})
\en
which integrates to
\eq
\phi_{b,L} = -\sqrt{\frac{2}{3 \kappa}} \log \sf + \frac{9A}{4}
\sqrt{\frac{2}{3 \kappa}} \sf^{4/3} +
\frac{V_{,\phi}}{4} \sf^2 + c_0+  O(\sf^{8/3})
\en
These are solutions to the second equation of motion (\ref{eom}),
up to $O(\ln \sf \sf^{2/3})$, provided
$A C^2 = 9/14$ and $V_{,\phi}$ small.

The background metric and field $\phi_b$ are to be matched at
equal values of the metric.   The approximate solution near
$\sigma \sim \sigma_f$ is only valid very close to
$\sigma_f$, so the matching will be done very close to
$\sigma \sim \sigma_f.$  Setting the scale factors and their
first derivatives equal,
\eq
\begin{array}{rll}
b_R(\sigma_m) &=& H^{-1}\sin H \sigma_m = C(\sigma_f-\sigma_m - \delta)^{1/3}
= b_L(\sigma_m + \delta)
\\
\nonumber
\part_\sigma b_R &=&
\cos H \sigma_m = -\frac{C}{3}(\sigma_f-\sigma_m - \delta)^{-2/3}
= \part_\sigma b_L
\label{bkgdm}
\end{array}
\en
If, for example,
$H \sigma_m \sim \pi$, neglecting the second term in $b_L$
($A\sfd^{4/3}$, down by 1/14) gives
\eq
\begin{array}{lll}
\frac{C}{3} &=& (\sigma_f-\sigma_m - \delta)^{2/3} \; \; \ll 1
\\
\nonumber
3 (\sigma_f-\sigma_m - \delta) &=& (\frac{\pi}{H} - \sigma_m) \; .
\label{approxm}
\end{array}
\en

For the background field values, using the solution in \cite{bgt} for
$\phi_{b,R}$,
\beqn
\phi_{b,R}(\sigma_m) &=& \phi_0  +
\frac{V_{,\phi}}{6 H^2}\left(-4 \ln \cos \frac{H \sigma_m}{2} +
\frac{1}{(\cos \frac{H \sigma_m}{2})^2} \right)\\
\nonumber \phi_{b,L}(\sigma_m + \delta)&=&
-\sqrt{\frac{2}{3 \kappa}} \log \sfd \\
\nonumber & &
+ \frac{3 A}{2} \sqrt{\frac{3}{2 \kappa}}\sfd^{4/3}
-\frac{V_{,\phi}}{4} \sfd^2 + c_0
\\
\nonumber
\part_\sigma
\phi_{b,R}(\sigma_m) &=& \frac{V_{,\phi}}{H}
\frac{\cos^3 H\sigma_m - 3 \cos H \sigma_m +2}
{3 \sin^3 H \sigma_m}
=
\frac{V_{,\phi}}{H}
\frac{2 + \cos H \sigma_m}{6 \cos^3 \frac{H \sigma_m}{2}}
\sin \frac{H \sigma_m}{2} \\
\nonumber &\sim_{H \sigma_m \sim \pi}
& \frac{V_{,\phi}}{H} \frac{4}{3 (H \sigma_m - \pi)^3}
\\
\nonumber
\part_\sigma \phi_{b,L}(\sigma_m + \delta)
&=&
\sqrt{\frac{2}{3 \kappa}}(\frac{1}{\sfd}- 3 A \sfd^{1/3}) + O(\sfd)
\label{bkgdphi}
\eeqn
Setting the first two lines equal can be used to determine
$\phi_0$, and setting the next two lines equal can be used to
determine $V_{,\phi}$.

Both of these are good approximate solutions for a potential of
the form
\eq
V = V_{,\phi} \phi
\en
with $V_{,\phi}$ small.
The numerical solutions shown in \cite{linde}
seem to have this approximate behavior as well.

\section{Scalar perturbations}
In open universe models where the fields tunnel from
a false vacuum, the tunneling `time' is on a proper Cauchy
surface exterior (except for a point, $\sigma = 0$) to the subsequent
open universe.  As
pointed out in \cite{gmst}, although this fixed time surface is a proper
Cauchy surface for quantization it is not a spatially homogeneous
background.  Therefore quantization in this background does
not split up into tensor, vector and scalar in the same way
as it does once the system is continued into the open universe.
For the calculations here, the distinction into scalar/vector/tensor
will be made on the basis of what the modes continue into in the open
universe.

At the fixed time of tunneling,
the background ($b(\sigma),\phi_b$) depends only on
$\sigma$.  In particular, $\phi_b$ is not a constant, unlike
many simpler cases ('thin wall' for example), and so the
scalar field and scalar metric perturbations become coupled.
It is better in this case to
use the `gauge invariant gravitational potential'
in the formalism of \cite{mukh-bard,bfm}, in particular its open
universe generalization detailed in section 7 of \cite{bgt}.
Equivalent descriptions using gauge fixing outside the
open universe are detailed in \cite{gmst}.

In the forward light cone (inside the open universe)
the gauge invariant potential $\Phi$ obeys \cite{bgt}
\eq
\ddot{\Phi} +
(\frac{\dot{a}}{a} - 2 \frac{\ddot{\phi}_b}{\dot{\phi_b}})
\dot{\Phi}
+\left(\frac{1}{a^2}(-\nabla^2 - 4 K)
+2 (\frac{\dot{a}}{a})^{\dot{}} -
2 \frac{\dot{a}}{a}\frac{\ddot{\phi}_b}{\dot{\phi}_b})\right)
\Phi = 0
\en

Continuing out of the light cone, and writing
$\dot{\Phi}=-i \part_\sigma \Phi$,
$\frac{\dot{a}}{a} = - i \part_\sigma b/b$
and $b^2 = - a^2$, this equation becomes
\eq
\part_\sigma^2\Phi +
(\frac{\part_\sigma b}{b} - 2 \frac{\part_\sigma^2{\phi_b}}{\part_\sigma\phi})
\part_\sigma\Phi
+(\frac{-\nabla^2 - 4K}{b^2}
+2 \part_\sigma \frac{\part_\sigma b}{b} -
2 \frac{\part_\sigma b}{b}
\frac{\part_\sigma^2\phi_b}{\part_\sigma\phi_b})\Phi = 0
\label{phieom}
\en

The gauge invariant gravitational potential can be expanded in terms
of basis functions
\eq
\Phi(\sigma,\tau,\theta,\phi) = \Phi_p (\sigma)Y_{p \ell m}(\tau,\theta,\phi)
\en
where the eigenfunctions of the three dimensional laplacian $Y_{p \ell m}$
obey
\eq
\nabla^2 Y_{p \ell m} = -(p^2 +1) Y_{p \ell m}
\en
and for an open universe $K = -1$.
The time dependence in $Y_{p \ell m}$ implicitly corresponds to
a particular
a definition of positive and negative frequency.
The surface where analytic continuation from the Euclidean to
Lorentzian signature is done defines a complex structure, as
described in, for example, \cite{gib}.  This method applies
rigorously when the surface of analytic continuation
is the only boundary of the space, so for the case here,
due to the singularity in
the instanton, this prescription is only a guide.
In terms of the five
dimensional embedding space, as described in \cite{yts},
at the surface of analytic continuation
$ \tilde{x}^0 = \sin \sigma \sin (i\tau) \to
i x^0 = \sin\sigma \sinh \tau $.
The Bunch Davies vacuum positive frequency wavefunctions are
regular for $\tilde{x^0} >0$ \cite{yts}.  For the tunneling prescription
\cite{gib} the wavefunctions must be bounded for $\tau_E = - i\tau <0$.
As $0 \le \sigma \le \pi$, $\sin \sigma \ge 0$ and hence
these two definitions of the split between positive and
negative frequency coincide.
One could also argue that since the O(3,1) symmetry persists in the presence of
this instanton \cite{garriga}, the Bunch Davies vacuum is appropriate for the
$\tau$ dependence.

In view of these arguments, the Bunch Davies vacuum will be used
for the $\tau$ dependent part of the fluctuations, and thus
the appropriate expression for $Y_{p \ell m}$
can be found for example in \cite{yts}:
\eq
Y_{p \ell m} \propto
\frac{P^{-\ell - 1/2}_{ip - 1/2}(\cosh(\tau + \pi i/2))}{
\sqrt{\sinh (\tau + \pi i /2)}} Y_{\ell m}(\theta,\phi)
\en
The $Y_{p \ell m}$ are normalized identically to earlier calcuations
such as those in \cite{ytss,jdc}, as all the background
$b(\sigma),\phi_b(\sigma)$ dependence affects only
$\Phi_p(\sigma)$.

We will want to solve equation \ref{phieom} for the
gauge invariant gravitational potential $\Phi$,
in the two asymptotic regimes
above.  Then these will be matched at the boundary between
the two backgrounds, given by equations \ref{bkgdm}, \ref{bkgdphi}.
The resulting basis function will then be normalized.

For the background $b_R(\sigma) = H^{-1}\sin H \sigma$, it is
convenient to use the conformal coordinate $u$,
defined by
\eq
\tanh u = \cos H \sigma\;, \;  \sin H\sigma = \frac{1}{\cosh u}\;, \;
\tan \frac{H\sigma}{2} = e^{-u}
\en
so $\part_\sigma = - H\cosh u \part_u $.
The two independent solutions to \ref{phieom} in this case
were found in \cite{bgt}
\beqn
\Phi_{R,+p}(u) &=& (e^{-u})^{1+ip}(1 -\frac{1-ip}{1+ip}
\frac{e^{-2 u}}{3})
\\
\nonumber
\Phi_{R,-p}(u) &=&
(e^{-u})^{1-ip}(1 - \frac{1+ip}{1-ip} \frac{e^{-2 u}}{3})
= \Phi_{R,+p}(u)^*
\eeqn

For the background with $b_L = C \sf^{1/3} (1 + A \sf^{4/3})$,
and
\eq
\part_\sigma \phi_{b,L} =
\sqrt{\frac{2}{3\kappa}}(\frac{1}{\sf} - 3 A \sf^{1/3}) - V_{,\phi}
\sf/2 + O(\sf^{5/3})
\en
The terms needed for the equation for $\Phi_p$ are
\eq
\frac{\part_\sigma b}{b} = \frac{-1}{3} \frac{\part_\sigma^2
\phi_b}{\part_\sigma \phi_b}
=-\frac{1}{3 \sf}(1 + 4 A \sf^{4/3} - \sqrt{\frac{3\kappa}{2}} V_{,\phi}
\sf^2))
\en
and
\eq
\frac{\part_\sigma b}{b} \frac{\part_\sigma^2 \phi_b}{\part_\sigma \phi_b} =
\frac{-1}{3 \sf^2} -
\frac{8 A}{3 \sf^{2/3}} - \sqrt{6\kappa} V_{,\phi}  + O(\sf^{2/3})
\; .
\en
The equation of motion for $\Phi$, to leading order and
assuming $\sigma_f - \sigma, \sqrt{6 \kappa} V_{,\phi}$ small
then becomes
\eq
\part_\sigma^2\Phi_{L,p} 
-\frac{7}{3 \sf} \part_\sigma \Phi_{L,p}
+\frac{p^2 + 9}{C^2 \sf^{2/3}} \Phi_{L,p} = 0 \; .
\en
This has solutions (via mathematica)
\eq
\Phi(\sigma) = c_1
\frac{J_1[3 \sqrt{\frac{p^2 + 9}{C^2}}
\frac{(\sigma_f-\sigma)^{2/3}}{2}]}{(\sigma_f-\sigma)^{2/3}}
+
c_2
\frac{K_1[3 i \sqrt{\frac{p^2 +9}{C^2}}
\frac{(\sigma_f-\sigma)^{2/3}}{2}]}{(\sigma_f-\sigma)^{2/3}}
\en
In terms of the coordinate
$X = \frac{3}{2}\sf^{2/3}$
the solution is
\eq
\Phi_{L,p}(X) = c_1
\frac{J_1[X \frac{\sqrt{p^2 + 9}}{C}]}{X}
+
c_2
\frac{K_1[iX \frac{\sqrt{p^2 + 9}}{C}]}{X}
\en
The asymptotics of the Bessel functions for small argument are
\eq
J_\nu(z) \sim \frac{(z/2)^\nu}{\Gamma(\nu + 1)} \; , \; \; \;
K_\nu(z) \sim \frac{1}{2} \Gamma(\nu) (\frac{2}{z})^\nu
\en
and so as $\sigma \to \sigma_f$,
\eq
\Phi_{L,p}(X) \sim c_1 \frac{\sqrt{p^2 +9}}{C}  +
c_2 \frac{C}{\sqrt{p^2 +9}}X^{-2} \; .
\en
The measure in the normalization, equation (\ref{phinorm}), goes as
$X^{3}$, and so the norm diverges for $c_2 \ne 0$.  Thus
we take $c_2 \to 0$, as mentioned in \cite{garriga}.
As $J_\nu(z e^{m \pi i}) = e^{m \pi \nu i} J_\nu(z)$, taking
the opposite sign in the argument of the Bessel function (the
square root of $p^2 + 9$) is
not linearly independent and consequently there is only one basis function
normalizable at the boundary.

The gauge invariant gravitational potentials
$\{ \Phi_{L,p}, \Phi_{R,\pm p} \}$ are now matched at
\eq
\sigma_L = \sigma_m + \delta \; , \; \; \sigma_R = \sigma_m
\en
with the backgrounds for $b, \phi_b$ given in
equations (\ref{bkgdm}, \ref{bkgdphi}).
The gravitational potentials
$\Phi_{L,p}(\sigma+ \delta)$ and
$\Phi_R(\sigma)$ must be matched at the same value of $|p|$ so that the
$\tau$ dependence agrees.
We have
\beqn
\Phi_{L,p}(\sigma_m+\delta) &=& \alpha_p \Phi_{R,p}(\sigma_m)
+ \beta_p \Phi_{R,-p}(\sigma_m) \\
\nonumber
\part_\sigma \Phi_{L,p}(\sigma)\vert_{\sigma =\sigma_m + \delta} &=&
\alpha_p \part_\sigma\Phi_{R,p}(\sigma)\vert_{\sigma = \sigma_m + \delta}
+ \beta_p \part_\sigma\Phi_{R,-p}(\sigma)\vert_{\sigma=\sigma_m + \delta}
\eeqn
Using
\eq
\part_\sigma = -H \cosh u \part_u
=- \sqrt{\frac{3}{2}}\frac{1}{X^{1/2}} \part_X
\en
it is also convenient to define
\eq
\tanh \tilde{u} = \cos H \sigma_m \; , \; \; \; \tilde{X} =
\frac{3}{2}\sfd^{2/3} \; .
\en
By the matching conditions of the background metric (\ref{bkgdm}),
we also have that
\eq
H \cosh \tilde{u} = \frac{1}{C}\sqrt{\frac{3}{2}}\frac{1}{\tilde{X}^{1/2}}
\en
The matching conditions become
\eq
\begin{array}{cl}
\frac{2}{3}\frac{J_1(\tilde{X} \frac{\sqrt{p^2 +9}}{C})}{\tilde{X}}
&= \alpha_p  (e^{-\tilde{u}})^{1+ip}(1 -\frac{1-ip}{1+ip}
\frac{e^{-2 \tilde{u}}}{3}) +
\beta_p (e^{-\tilde{u}})^{1-ip}(1 -\frac{1+ip}{1-ip}
\frac{e^{-2 \tilde{u}}}{3})
\\
\nonumber
- \sqrt{\frac{3}{2}}\frac{1}{X^{1/2}} \part_X
\frac{2}{3}\frac{J_1(X \frac{\sqrt{p^2 +9}}{C})}{X}
_{X = \tilde{X}}
&=
-\alpha_p H \cosh u \part_u  \left(e^{-u})^{1+ip}(1 -\frac{1-ip}{1+ip}
\frac{e^{-2 u}}{3}\right)_{u = \tilde{u}} \\
&\; \; \; \; \;
 -\beta_p H \cosh u \part_u  \left(e^{-u})^{1-ip}(1 -\frac{1+ip}{1-ip}
\frac{e^{-2 u}}{3}\right)_{u = \tilde{u}}
\end{array}
\en
and so, with
\beqn
\alpha_p &=& \frac{\part_\sigma \Phi_{R,-p} \Phi_{L,p} -
\Phi_{R,-p} \part_\sigma \Phi_{L,p}}
{\part_\sigma \Phi_{R,-p} \Phi_{R,p} -
\Phi_{R,-p} \part_\sigma\Phi_{R,p}}
\\
\nonumber
\beta_p &=&\frac{- \part_\sigma \Phi_{R,p} \Phi_{L,p} +
\Phi_{R,p} \part_\sigma \Phi_{L,p}}
{\part_\sigma \Phi_{R,-p} \Phi_{R,p} -
\Phi_{R,-p} \part_\sigma\Phi_{R,p}}
\eeqn
we get
\beqn
\alpha_p &=&
e^{\tilde{u}(1 + ip)}
\frac{
(-1 + ip -\frac{e^{-2\tilde{u}}}{3} \frac{1 + ip}{1-ip}(-3 + ip))
\frac{J_1(\tilde{X} \frac{\sqrt{p^2 +9}}{C})}{\tilde{X}}
- C(1-\frac{1+ip}{1-ip}\frac{e^{-2\tilde{u}}}{3})\part_{\tilde{X}}
\frac{J_1(\tilde{X} \frac{\sqrt{p^2 +9}}{C})}{\tilde{X}}}
{3 ip(1+\frac{e^{-2\tilde{u}}}{3})^2} \\
\nonumber &=&
e^{\tilde{u}(1 + ip)}
\frac{
(-1 + ip - 2 \tanh \tilde{u}-
\frac{e^{-2\tilde{u}}}{3} \frac{1 + ip}{1-ip}(-3 + ip- 2 \tanh \tilde{u}))
J_1(y)
- (1-\frac{1+ip}{1-ip}\frac{e^{-2\tilde{u}}}{3})
\frac{\sqrt{p^2 + 9}}{2}
[J_0(y)-
J_2(y)]}
{3 ip \tilde{X}(1+\frac{e^{-2\tilde{u}}}{3})^2} \\
\nonumber y &=& \frac{-\sqrt{p^2 + 9}}{2 \tanh \tilde{u}}
\label{alphap}
\eeqn
The second line comes from using the matching conditions
equation (\ref{bkgdm}), i.e. that $C = - 2 \tanh \tilde{u} \tilde{X}$.
Also,
\eq
\beta_p = \alpha_p^*
\en
which could be seen as well from noting that $\phi_L$ is real.

Thus the unnormalized gauge invariant gravitational potential
$\Phi_p$ is
\beqn
\Phi_p &=& \Phi_L(\sigma) \; \; , \; \; \sigma_f > \sigma > \sigma_m + \delta
\\
\nonumber
&=& \alpha_p \Phi_{R,+} + \alpha_p^* \Phi_{R,-} \; \; , \; \; \;
\sigma_m > \sigma \ge 0 \\
\eeqn

The measure for normalizing $\Phi_p$ comes from
relating $\Phi_p$ to the scalar field fluctuation.
The Klein Gordon normalization for the scalar field
translates into \cite{bfm,gmst}
\eq
\int d \sigma \, \frac{1}{b(p^2 +4)} (\frac{2}{\kappa \part_\sigma \phi_b})^2
\Phi_p \Phi_{p^\prime} = \delta (p - p^\prime) N_p^2
\label{phinorm}
\en
Thus
\beqn
(p^2 +4)\delta(p-p^\prime)N_p^2 &= &\int_0^{\sigma_f}\frac{d\sigma}{b(\sigma)}
(\frac{2}{\kappa \part_\sigma \phi_b})^2\Phi_p(\sigma)\Phi_{p^\prime}(\sigma)
\\
\nonumber
&=&
\int_0^{\sigma_m}\frac{d\sigma}{b(\sigma)}
(\frac{2}{\kappa \part_\sigma \phi_{R,b}})^2
4 \Re(\alpha_p \Phi_{R,+p}(\sigma))
\Re(\alpha_{p^\prime} \Phi_{R,+p^\prime}(\sigma))
\\
\nonumber
&+&
\int_{\sigma_m + \delta}^{\sigma_f}\frac{d\sigma}{b(\sigma)}
(\frac{2}{\kappa \part_\sigma \phi_b})^2\Phi_{L,p}(\sigma)
\Phi_{L,p^\prime}(\sigma)
\eeqn
As an eigenfunction of a self-adjoint operator, the rescaled $\Phi_p$ is
orthogonal to any eigenfunction with a different value of
$p^2$.  Thus one can read off $N_p^2$
from the behavior at the endpoints
(although there is a jump in $\sigma$ in the integrand, continuity
in the argument is ensured by the matching conditions).
On the other hand, as $\sigma \to 0$ (or $u \to \infty$),
\eq
\Phi_p \sim \alpha_p e^{-u(1+ip)} + \alpha_p^* e^{-u(1-ip)}
\label{abasis}
\en
and
\eq
\part_\sigma \phi_R \sim
\frac{V_{,\phi}}{2 H}e^{-u} \; .
\en
The asymptotics of the inner product in equation \ref{phinorm}
thus give
\eq
\begin{array}{ll}
\int du (\frac{4}{\kappa})^2\frac{4}{p^2 + 4} (\frac{H}{V_{,\phi}})^2 &
(\alpha_p e^{-ipu} + \alpha_p^* e^{ipu})
(\alpha_{p^\prime} e^{-ip^\prime u} + \alpha_{p^\prime}^* e^{ip'u}) \\
\nonumber
&\sim
 \pi (\frac{4}{\kappa})^2
\frac{4}{p^2 + 4} (\frac{H}{V_{,\phi}})^2
\left[
(\alpha_p \alpha_{p^\prime} + \alpha_p^* \alpha_{p^\prime}^*)
\delta(p+p^\prime) +
(\alpha_p \alpha_{p^\prime}^* + \alpha_p^* \alpha_{p^\prime})
\delta(p-p^\prime) \right]
\end{array}
\en
This is symmetric under $p \to - p$ just as the basis functions are,
so 
restricting to $p>0$ the inner product becomes
\eq
\alpha_p \alpha_p^* (\frac{4}{\kappa})^2
\frac{8 \pi}{p^2 + 4} (\frac{H}{V_{,\phi}})^2
 = N_p^2
\en

This gauge invariant gravitational potential can then be continued
into the forward light cone for the open universe, and then used to
compute a power spectrum,
\eq
\langle \Phi(\chi=0,t) \Phi(\chi,t) \rangle
= \int_0^\infty dp \, p \frac{\sin p \chi}{\sinh \chi} P_\Phi(p,t) \; .
\en
As the basis functions used here are the same as those in earlier
calculations\cite{bgt,ytss,jdc}, expressions for
$P_\Phi$ can be used to complete the calculation.
The normalization $N_p^2$ found here
corresponds to taking $b_+ = 
\frac{1}{\vert \alpha_p \vert}$
and $b_- = 0$ in expression (5.9) of 
\cite{jdc}, and not summing over both $\pm p$.  
As a result,
\eq
P_{\Phi(p, t \to \infty)}  = \frac{(G V_{,\phi})^2}
{ p(p^2 +1)2 \sinh \pi p}\frac{4}{9H^2}
\vert e^{\pi p/2}  \frac{\alpha_p}{\vert \alpha_p \vert}
+\frac{1-ip}{1 + ip}\frac{\alpha_p^*}{\vert\alpha_p \vert}e^{-\pi p/2}\vert^2
\label{phip}
\en
Defining $\alpha_p = A_p e^{i \lambda_p}$ with $A_p, \lambda_p$ real,
\eq
P_\Phi =
\frac{4}{9 H^2}
(GV_{,\phi})^2 \frac{(e^{\pi p} + e^{-\pi p}-
2 \frac{p^2-1}{p^2 +1} \cos 2 \lambda_p
-\frac{2p}{p^2 +1} \sin 2 \lambda_p)}{p(1+p^2)2 \sinh \pi p}
\en
We see that just as for the subcurvature modes in the
presence of a bubble wall, for $p \ge 1$ the last 3 terms are bounded
and become irrelevant.

\section{Supercurvature modes}
Supercurvature modes may appear in a field's expansion in
an open universes\cite{lw,yts,mos}. 
At first it was unclear whether they appeared in the vacuum expansion of
a quantum field, but calculations \cite{yts} (see also \cite{mos}) 
of the wightman function showed that they were
required in the sum over modes in order to produce the
Bunch Davies vacuum.
Their effects can be large, and thus in some cases they 
constrain models of open inflation
\cite{st-can}.  
A supercurvature mode will have $p$ imaginary.  Take
$-ip = \Lambda >0$, without loss of generality.
In this case, the asymptotics of the basis function at
$u \sim \infty$, equation \ref{abasis} requires
$\alpha_{i\Lambda} = 0$.
Normalizability at $\sigma \sim \sigma_f$ does not seem to constrain the
the value of $p$ as the basis function is bounded there
independent of $p$.

One can look for a supercurvature mode in the case
where $H \sigma_m - \pi$ small,
i.e. where equation (\ref{approxm}) holds. Then
$\tilde{X}/C = \frac{1}{2}$ and the equation (\ref{alphap}) for
$\alpha_{p = i \Lambda}$ becomes
\beqn
\alpha_{i \Lambda}
&  \propto &
(-1  -\Lambda +\frac{e^{-2\tilde{u}}}{3} \frac{1 -\Lambda}{1 +\Lambda}
(3 + \Lambda))\frac{J_1(\frac{\sqrt{9- \Lambda^2}}{2})}{\tilde{X}} \\
\nonumber
& &- (1-\frac{1-\Lambda}{1+\Lambda}\frac{e^{-2\tilde{u}}}{3})
\frac{\sqrt{9 - \Lambda^2} (J_0 (\jarg)- J_2(\jarg))
-4 J_1(\jarg)}{2 \tilde{X}}
\\ & \equiv & 0
\eeqn
The $\tilde{X}$ dependence thus falls out and we need to solve
\eq
0 =
(-1  -\Lambda +\frac{e^{-2\tilde{u}}}{3} \frac{1 -\Lambda}{1 +\Lambda}
(3 + \Lambda))J_1(\frac{\sqrt{9- \Lambda^2}}{2})
- (1-\frac{1-\Lambda}{1+\Lambda}\frac{e^{-2\tilde{u}}}{3})
\frac{\sqrt{9 - \Lambda^2} (J_0 (\jarg)- J_2(\jarg))
-4 J_1(\jarg)}{2}
\en
This is satisfied for a range of $e^{-2 \tilde{u}}$ values
(corresponding to a choice of model), as can be seen by
plotting the right hand side.
The approximations in equation (\ref{approxm}) require $e^{-2\tilde{u}}$
to be large enough that $\tanh \tilde{u} \sim -1-2 e^{2 \tilde{u}} \sim -1$.
Taking $e^{2 \tilde{u}} = .01$, this normalizability
equation for the supercurvature
mode has a solution for $\Lambda \sim .97$.  Increasing the
value of $e^{-2\tilde{u}} $ increases the required value of $\Lambda$.
Thus it seems that a normalizable supercurvature mode may be present.
The interpretation for $\Lambda >1$, where the eigenvalue
of the spatial laplacian $-(p^2 +1)$ changes sign, seems unclear.

\section{Gravity waves}
For gravity waves, again the $\tau,\theta,\phi$ dependence of the
gravity waves in this instanton background is the same as the earlier
calculations in \cite{st,bc,gmst}.
The gravity waves are perturbations on the background metric
\eq
g_{\mu \nu} = g_{\mu \nu}^b + \hat{h}_{\mu \nu}
\en
where $\hat{h}$ is transverse and traceless.
Expanding $\hat{h}^i_j$ in terms of tensor eigenfunctions
${\bf T^{P,p \ell m}}^i_j (\tau,\theta,\phi)$ of
the $\tau,\theta,\phi$ laplacian, one gets an expansion in terms
of modes
\eq
n(p) {\bf T^{P,p \ell m}}^i_j (\tau,\theta,\phi) T_p(\sigma)
\en
The instanton behavior appears in the $\sigma$ dependence, and
$T_p$ has the
equation of motion
\eq
\part_\sigma^2 T_p + 3 \frac{\part_\sigma b}{b} \part_\sigma T_p +
\frac{p^2 +1}{b^2} T_p = 0.
\label{theq}
\en

The normalization \cite{gmst} is proportional to
\eq
\int d \sigma b^3(\sigma) \part_\sigma(T_p) \part_\sigma (T_p^*)
=
(p^2 +1)\int d\sigma b (\sigma) T_p T_p^*
\label{garnorm}
\en

For $b_R = \sim H^{-1} \sin H \sigma = \frac{1}{H \cosh u}$,
the fluctuations are
\beqn
T_{R,+p}(u) &=& (\tanh u - ip )\cosh u e^{i p u} \\
\nonumber
T_{R,-p}(u) &=& (\tanh u + ip) \cosh u e^{-ipu}
\eeqn
while for
$b_L = C \sf^{1/3}= C \sqrt{\frac{2}{3} X}$,
\eq
T_{L,p}(\sigma) =
c_1 J_0[ \frac{\sqrt{1+p^2}}{C}X]
+ c_2 K_0 [i \frac{ \sqrt{1+p^2}}{C} X]
\en
Since
\beqn
J_0(x) & \sim &  1 + \frac{x^2}{4} + O(x^2) \;
\\ \nonumber
K_0(x) &\sim & (k_1 - \ln x) +
(k_2 - \frac{\ln x}{4})x^2 + O(x^3)
\eeqn
with $k_1,k_2$ constants, one can see
using eqn.(\ref{garnorm}) that normalizability requires
$c_2= 0$.
Thus,
\eq
T_{L,p}(X)= J_0(\frac{\sqrt{p^2 +1}}{C}X)
\en
Matching these two asymptotic values and their derivatives at the
juncture given by equations (\ref{bkgdm}, \ref{bkgdphi}),
\eq
\begin{array}{l}
J_0(\frac{\sqrt{p^2 +1}}{C}\tilde{X}) =
\alpha_{h,p} (\tanh \tilde{u} - ip )\cosh \tilde{u} e^{i p \tilde{u}}
+ \beta_{h,p} (\tanh \tilde{u} +ip )\cosh \tilde{u} e^{-i p \tilde{u}} \\
-\sqrt{\frac{3}{2 \tilde{X}}}
\part_X J_0(\frac{\sqrt{p^2 +1}{C}}X)|_{X = \tilde{X}}
=
-\alpha_{h,p}
H\cosh^2\tilde{u} (1+p^2)e^{ip\tilde{u}} -
\beta_{h,p} H\cosh^2 \tilde{u} (1+p^2) e^{-ip\tilde{u}} \; .
\end{array}
\en
So,
\beqn
\alpha_{h,p} &=&
\frac{(1+p^2)\cosh^2\tilde{u}  J_0(\frac{\sqrt{p^2 +1}}{C}\tilde{X})-
 C \part_X J_0(\frac{\sqrt{p^2 +1}}{C}\tilde{X})
(\sinh \tilde{u} + ip \cosh \tilde{u})  }
{-2 i p\cosh^3 \tilde{u} (1+p^2)} e^{-ip\tilde{u}} \\
\nonumber &=&
\frac{(1+p^2)\cosh^2\tilde{u}  J_0(y_h)
-\frac{\sqrt{p^2 +1}}{2}[J_{-1}(y_h) - J_1(y_h)]
(\sinh \tilde{u} + ip \cosh \tilde{u})  }
{-2 i p\cosh^3 \tilde{u} (1+p^2)} e^{-ip\tilde{u}}
\\
\nonumber y_h &=& \frac{-\sqrt{p^2 +1}}{2 \tanh\tilde{u}}
\label{alphp}
\eeqn
and again
\eq
\beta_{h,p} = \alpha_{h,p}^* \; .
\en
Equation \ref{garnorm} can be used to get the normalization
dependence on $\alpha_p,\beta_p$, which is proportional to
$|\alpha_p|^2$.  The power spectrum
can be taken over from the results in \cite{st,bc}.
The two point function dependence on $T_{h,p}$ goes as
\eq
P_{h}(p)\sim H^2 \frac{1}{2p(p^2 +1)\sinh \pi p}
|e^{\pi p/2} \frac{\alpha_{h,p}}{\vert \alpha_{h,p} \vert} +
e^{-\pi p/2} \frac{\alpha_{h,p}^*}{\vert \alpha_{h,p} \vert}|^2
=H^2 \frac{e^{\pi p} + e^{- \pi p} +
2 \frac{\alpha_{h,p}^2 + \alpha_{h,p}^{2 \, *}}{\vert \alpha_{h,p} \vert^2}}
{2p(p^2 +1) \sinh \pi p}
\label{powerh}
\en
times $p$ independent terms.
Writing $\alpha_{h,p} = A_{h,p} e^{i \lambda_{h,p}}$,
equation \ref{alphp} implies that
\eq
\lambda_{h,p} = \frac{-\pi }{2} + O(p)
\en
so that as $p \to 0$, equation (\ref{powerh}) goes as
\eq
\frac{e^{\pi p} + e^{- \pi p} +
2 \cos \lambda_{h,p}}{2p(p^2 +1) \sinh \pi p}
\rightarrow_{p \to 0}
\frac{2 + 2 \cos(-\pi + O(p))}{p^2}
= \frac{O(p^2)}{p^2} = O(1)
\en
which is finite.  As the $p \to 0$ divergence in the
pure Bunch Davies vacuum \cite{all-cal} does not appear,
the gravity wave contribution to the CMB is finite.

One can also look for the possibility of normalizable supercurvature
modes here, although they are not present in other cases
which have been studied so far \cite{st}.  Setting
$ip = \Lambda$ in equation (\ref{alphp}),
and using the approximation equation (\ref{approxm}), it
seems that one cannot have normalizability, i.e.
$\alpha_{h,-i\Lambda} = 0$, for $0 \le \Lambda \le 1$.

\section{Conclusion}
Assuming the background model
and resulting $\Omega$ value are consistent,
subcurvature scalar and gravity wave fluctuations were found
for a background approximating that described in \cite{ht}.
The initial conditions
were found by requiring regularity at the singular boundary along
with the time dependence (on the surface of analytic
continuation) corresponding to the Bunch Davies
vacuum.  Even though a bubble
wall is not present, the gravity waves were found to be finite.
Both scalar and gravity wave power spectra, for `momentum' $p \ge 2$,
go over to the conformal vacuum case (found for scalars in
\cite{cv}).
A normalizable scalar supercurvature mode appears
possible for some values of the matching parameters and
might have observable effects.
The power spectra for $P_\Phi$ and $P_{h}$ can be used to
calculate cosmic microwave background anisotropies.  The
the subcurvature scalar power spectrum $P_\Phi$ lies within the same envelope
as subcurvature modes in other open universe models.   
Thus varying parameters in this background will
change the subcurvature scalar contribution to the
CMB in a bounded way and  
only affect the largest scales just as in the other 
open universe models. 
The gravity wave contribution to the CMB is finite.  
The corresponding gravity wave CMB spectrum
may provide constraints on viable backgrounds, just as
it does in some of the open universe models (see, e.g. \cite{gw-const}).

\section{Acknowledgements}
This work was supported by an NSF Career Advancement Award,
NSF-PHY-97-22787.
I thank S. Carlip and M. White for discussions, and J. Garriga for
comments on the draft.

\end{document}